\documentstyle[epsf,a4wide,12pt]{article}
\begin{document}
\baselineskip=18pt
\begin{tabbing}
\hskip 11.5 cm \= {Imperial/TP/94-95/33}\\
\>hep-th/9505099\\
\> Submitted to {\em Nuclear Physics B}\\
\>May 1995
\end{tabbing}
\vskip 1cm
\begin{center}
{\Large\bf Higher-loop anomalies in chiral gravities}
\vskip 1.2cm
{\large \bf R. Mohayaee$^a$ and L. Wenham$^b$}\\
{{\it The Blackett Laboratory, Imperial College, Prince Consort Road\\
 London SW7 2BZ  U.K.}\\
$^a$ r.mohayaee@ic.ac.uk\\
$^b$ wenham@ic.ac.uk}
\end{center}

\abstract
{The one-loop anomalies for chiral $W_{3}$ gravity are derived using the
Fujikawa regularisation method. The expected two-loop anomalies are then
obtained by imposing the
Wess-Zumino consistency conditions on the one-loop
results. The anomalies found in this way agree
with those already known from explicit Feynman diagram
calculations. We then directly verify that
the order $\hbar^2$ non-local BRST Ward identity anomalies, arising
from the ``dressing'' of the one-loop results,
satisfy Lam's theorem. It is also shown that in a rigorous calculation of
$Q^2$ anomaly for the BRST charge, one recovers both the non-local as well as
the local anomalies. We further verify
that, in chiral gravities, the non-local anomalies in
the BRST Ward identity
can be obtained by the application of
the anomalous operator $Q^2$,
calculated using operator products, to an appropriately defined
gauge fermion. Finally,
we give arguments to show why this relation should hold generally in
reparametrisation-invariant theories.}

\vfill\eject
\section{Introduction}
\indent

The r\^oles that anomalies take in different theories can vary
substantially
according to the physical content of a given theory.
The existence of anomalies in four-dimensional
theories can render them non-renormalizable and non-unitary, while
 in two dimensions, because of
the softer ultraviolet behaviour, anomalies may prove not to be disastrous.
Furthermore, the
renormalizability of two-dimensional theories, the relative
simplicity of the
forms of propagators and consequently of operator products and also the extent
to which the BRST formalism can be used, have helped to reveal
new features of anomalies.

There exist two basic approaches to the treatment of
anomalies. One standard approach is to anticipate
the occurrence of the anomalies already at the classical level by
introducing classically-decoupling compensating fields. A well-known
illustration of this would be when anomalies give rise to propagating
Liouville modes. An alternative
approach, to be adopted here, is to extract the
anomaly-induced dynamics directly from the Ward identities
of the worldsheet symmetries.

It is known that the two-dimensional ${\cal W}$
gravity theories have anomalies beyond the one-loop order. Anomalies
 occur at the two-loop order
for $W_3$ and at all higher-loop orders for $W_\infty$ and
$w_\infty$ gravities. The presence of such
higher-loop anomalies make these theories particularly interesting to
study. In this paper we investigate the anomalies in the fully
quantised chiral $W_3$ theory, considering specifically those appearing up to
order $\hbar^2$. This formalism is outlined in the following section.

It turns out that discussing anomalies beyond the one-loop order is
not simply a matter of extending existing methods. Some techniques employed
to evaluate one-loop anomalies are restricted in applicability to that
order alone.
For instance, in Section 3, the one-loop anomalies shall be calculated
using the path-integral method initiated by Fujikawa
\cite{fujikawa}\cite{diaz}. This is a purely one-loop
approach and it is
still unknown how the two-loop anomalies can arise
in such a path-integral formalism. In Section 4, we shall show that
this deficiency of the Fujikawa method can be overcome for our
purposes by using the Wess-Zumino consistency condition
to derive the two-loop anomalies.

A noteworthy distinction between the one- and two-loop anomalies is that,
while those at the one-loop order are necessarily local, those at
higher-loop orders are not. The non-local order $\hbar^2$
anomalies arising in our theory can be used to make a non-trivial verification
of Lam's theorem \cite{lam}. This theorem states that all the non-local
contributions to the Ward identity anomaly can be obtained via the
``dressing'' of lower order local anomalies. The explicit verification
of this is the subject of Section 5.

Anomalies can also be studied in the context of the BRST charge $Q$. The
non-nilpotence of the BRST charge at the full quantum level is another
expression of an anomaly. Although it is known how local anomalies
are obtained in the BRST anomalous algebra \cite{mohayaeestelle}, little
attention has been paid to the non-local anomalies. In Section 6, both
local and
non-local anomalies shall be derived from the expectation value
of the square of the BRST charge.

It has recently been noted that the BRST algebra
anomalies in chiral gravities can be related to those
of the BRST Ward identity via a
construction involving an appropriately defined
gauge fermion \cite{mohayaeestelle}.
Specifically, the
Ward-identity anomalies are obtained by the application of the
anomalous operator $Q^2$ to this gauge fermion. This relation has as
yet been discussed only for local anomaly contributions. In
Section 7, we shall show that the anomaly relation also holds for the
non-local terms. In Section 8, we shall show that this relationship
is valid for a more general gauge-fixing
condition. Lastly, Section 9 considers the general origin of this
anomaly relation.

\section{BRST Quantisation of $W_3$ gravity}
\indent

The classical action for chiral $W_3$ gravity is \cite{hull}
\begin{eqnarray}
I_{\rm class} &=& {1\over\pi}\int d^2 z\bigg(-{1\over 2}\bar\partial
\phi^i \partial\phi^i + {1\over 2}h\partial\phi^i \partial\phi^i
+ {1\over 3}B d_{ijk}\partial\phi^i \partial\phi^j \partial\phi^k
\bigg), \nonumber\\
& &
\end{eqnarray}
where $\phi^i\,  (i=1\cdots n)$ are scalar fields, $h$ and $B$ are the
spin-2 and spin-3 gauge fields respectively and $d_{ijk}$ is a
constant symmetric traceless tensor. Also, $\partial = \partial /
\partial z ,\bar\partial = \partial /
\partial \bar{z}$ and Einstein summation is assumed throughout.

This action is invariant under the following nonlinear gauge symmetries:
\begin{eqnarray}
\delta\phi^i &=& \epsilon\partial\phi^i +\nu d_{ijk}\partial\phi^j
\partial\phi^k, \nonumber\\
\delta h &=& \bar\partial\epsilon + \epsilon\partial h
-\partial\epsilon h - {1\over 2}(\nu\partial B -\partial\nu
B)a\partial\phi^i \partial\phi^i , \nonumber\\
\delta B &=& \bar\partial\nu + \epsilon\partial B -2\partial\epsilon B
+ 2\nu\partial h -\partial\nu h,
\end{eqnarray}
where $\epsilon$ is the spin-2, $\nu$ is the spin-3 infinitesimal
parameter and $a$ is a constant.

Imposing derivative gauge-fixing conditions,  $\bar\partial h=0$ and
$\bar\partial B=0$,  we necessarily extend the phase space to include
spin-2 Faddeev-Popov ghost and antighost fields $c, b$,  and, likewise, in the
spin-3 sector, $\gamma, \beta$.
The resulting BRST-transformations are not nilpotent off-shell
(i.e.\ without imposing the equations of
motion) \cite{schoutens}\cite{pope}. Learning a
lesson from supergravity \cite{kallosh},  we see that this difficulty
can be overcome
by the introduction of auxiliary fields. For $W_3$ gravity we find
that the momenta conjugate to the ghosts and anti-ghosts
naturally play the role of such
 auxiliary fields. Hence, we introduce
these momenta $\pi$ and reduce the action
to first-order form.\footnote{This nilpotency can equally
well be achieved in the
conformal gauge,  by eliminating the gauge fields from the
gauge-fixed action as seen in the Batalin-Vilkovisky formalism
 \cite{vandoren}.}
Finally,  requiring
renormalisation terms to achieve the cancellation of matter-dependent
anomalies,  we arrive at the extended quantum action of chiral $W_3$
gravity in the derivative gauge and in the first order
form \cite{mohayaeestelle},
\begin{eqnarray}
S&=&{1\over\pi}\int \bigg[-{1\over 2}
\bar\partial\phi^i \partial\phi^i+\pi_h\bar\partial h+\pi_B\bar\partial
B-\pi_b\bar\partial b-\pi_c\bar\partial
c-\pi_\beta\bar\partial\beta-\pi_\gamma\bar\partial\gamma\nonumber\\
& &
-\delta(h\pi_c+B\pi_\gamma) + K_{\varphi^i} \delta\varphi^{i}\bigg],
\end{eqnarray}
where $\delta$ indicates a BRST variation, $\varphi^{i}$ denotes all
fields and momenta and $ K_{\varphi^i} $ are the corresponding
sources for their variations.
 The action is expressed in this form for
brevity as well as to make the classical BRST invariance manifest.
A peculiarity of this ``canonical'' action, which derives from its origin
in a fully reparametrisation-invariant theory, is that the ``Hamiltonian''
$\delta(h\pi_c+B\pi_\gamma)$ is BRST exact. We shall see that this has
important consequences later.

The full BRST symmetries of the action are given by:
\begin{eqnarray}
\delta {\phi }^i &=&c\,  {\partial {\phi }^i} + d_{ijk}\,
\gamma\,
{\partial {\phi }^j} {\partial {\phi }^k} + a \, {\pi}_c\,  \gamma
\, {\partial
\gamma }\,  {\partial {\phi }^i}
+{\sqrt \hbar } \Big( -a\,  {\alpha }_i\,
\partial ({\pi }_c\,  \gamma\,
\partial \gamma )\nonumber\\
& & -{\alpha }_i\,
\partial c + (e_{ij}-e_{ji}) \, \gamma\,  {\partial }^2 {\varphi }^j -
e_{ji}\,
\partial \gamma \, \partial {\phi }^j \Big)+
\hbar\,  f_i \, {\partial }^2 \gamma , \nonumber\\
\delta h &=& {\pi}_b  ,  \qquad\nonumber\\
\delta B &=& {\pi}_{\beta} ,  \nonumber\\
\delta c &=& c \, \partial c -{a\over 2} \gamma \, \partial \gamma\,
\partial {\phi }^i \partial {\phi }^i - a {\sqrt \hbar } \, {\alpha
}_i\,
\gamma \, \partial \gamma \, {\partial }^2 {\phi }^i +{{1-17 a}\over 30}
\hbar ( 2\gamma \, {\partial }^3 \gamma - 3 \partial \gamma\,
{\partial }^2 \gamma ) ,  \nonumber\\
\delta \gamma &=& c \, \partial \gamma - 2 \partial c \, \gamma , \nonumber\\
\delta b &=& {\pi }_h  , \ \ \ \  \delta \beta = {\pi }_B  ,   \ \ \ \
\delta {\pi}_c = T_{\rm mat} + T_{\rm gh}
  , \ \ \ \  \delta {\pi}_{\gamma}=
W_{\rm mat} + W_{\rm gh} , \nonumber\\
\delta {\pi }_h &=& 0 ,
\ \ \ \  \delta {\pi}_B = 0 , \ \ \ \  \delta {\pi}_b = 0 ,
\ \ \ \  \delta {\pi }_{\beta } = 0, \nonumber\\
\delta {K_{\varphi^i}} &=& 0 ,
\end{eqnarray}
where the spin-2 and spin-3 matter currents are
\begin{eqnarray}
T_{\rm mat} &=& -{1\over 2}\partial {\phi}^i\partial {\phi}^i -{\sqrt \hbar}
{\alpha}_i {\partial}^3 {\phi}^i , \nonumber\\
W_{\rm mat} &=& -{1\over 3}d_{ijk}\partial {\phi}^i\partial
{\phi}^j\partial {\phi}^k
-{\sqrt \hbar} e_{ij}\partial {\phi}^i{\partial}^2 {\phi}^j
-\hbar f_i{\partial}^3 {\phi}^i ,
\end{eqnarray}
and the corresponding ghost currents are given by:
\begin{eqnarray}
T_{\rm gh} &=& -2{\pi}_c\, \partial c -\partial{\pi}_c\,  c -3{\pi}_{\gamma}
\, \partial\gamma
-2\partial\pi_{\gamma}\, \gamma , \nonumber\\
W_{\rm gh} &=& -\partial\pi_{\gamma}\,  c -3\pi_{\gamma}\, \partial c
-a[\partial (\pi_{c}\gamma T_{\rm mat}) + \pi_{c}\partial\gamma T_{\rm
mat}]\nonumber\\
&+& {(1-17a)\over 30}\hbar (2\gamma\, {\partial}^3 \pi_{c} +
9\partial\gamma\, {\partial}^2 \pi_{c} + 15{\partial}^2 \gamma\,
\partial\pi_{c}
+ 10{\partial}^3\gamma\,  \pi_{c}).\nonumber\\
& &
\end{eqnarray}

To insure the cancellation of all the matter-dependent anomalies that
arise in this theory, the symmetric
structure constant $d_{ijk}$, the background-charge
$\alpha_i$ and the renormalisation constants $e_{ij}$ and $f_i$
must satisfy the following conditions \cite{pope}:
\begin{eqnarray}
 & & d_{ijj} - 6 e_{ij}\,  {\alpha }_j + 6 f_i = 0 ,  \nonumber\\
& & e_{(ij)} - d_{ijk}\,  {\alpha }_k = 0 ,  \nonumber\\
& & 3 f_i - {\alpha }_j\,  e_{ji} = 0 ,  \nonumber\\
& & d_{ikl}\, d_{jkl} + 6 d_{ijk}\,  f_k -3 e_{ik}\,  e_{jk}
 = {1\over 2} {\delta }_{ij} , \nonumber\\
& & d_{(ij}{}^m\,  d_{kl)m} =
{1\over 2} a \, {\delta }_{(ij} {\delta }_{kl)} , \nonumber\\
& & d_{ijk} ( e_{lk} - e_{kl} ) + 2 e_{(i}{}^l\,  d_{j)kl} = a\,  {\alpha }_k
\, {\delta }_{ij} ,
\end{eqnarray}
where the constant $a$ is equal to $16/(22+5C_{\rm mat})$.\footnote
{Interestingly, these conditions are in fact
identical to those of Romans,  designed to obtain quantum $W_3$
symmetry \cite{romans}. This identification also holds in
$W_\infty$ \cite{bergshoeff}
although there is as yet no proof that it must hold generally.}
(Note that
at the quantum level, $d_{ijk}$ is no longer traceless ).
The central charge $C_{\rm mat}$ is given in terms of the number of
scalar fields and the background charge by the following
expression:
\begin{eqnarray}
C_{\rm mat} &=& n + 12 {\alpha }_i {\alpha }_i ,
\end{eqnarray}
or, equivalently, it can be shown that:
\begin{eqnarray}
C_{\rm mat}&=&2d_{ijk}\, d_{ijk}-18e_{ij}\, e_{ij}-12e_{ij}\, e_{ji}
-360f_i^2.
\end{eqnarray}

Note that, although usually one would expect in field theory that
the BRST variations of the original fields are just gauge transformations
with the gauge parameters replaced by the corresponding ghosts,  this
simple replacement fails for $W_3$ gravity.  The root of
this feature can be seen even at the level of the gauge
 transformation of the spin-2 gauge field
(2),  which transforms  non-linearly into the scalar fields. The original
gauge-invariant part of the action is not invariant under the BRST
symmetries.
To compensate for this non-invariance, one needs to modify the
transformation of the scalar fields $\phi$ by
 the additional term $a\pi_c\gamma\partial\gamma\partial\phi$.

\section{The derivative-gauge anomalies of $W_3$ gravity}
\indent

The presence of anomalies is expressed in the Ward identity,
which in turn results from requiring the invariance of the
partition function under the symmetries of the action.
The partition function corresponding to the action $S$ (3) is given by
a path integral over all fields and their conjugate momenta,
weighted by the exponential of the extended action, namely:
\begin{equation}
Z=\int{\cal D}\varphi^{i}\exp S_{\rm ext},
\end{equation}
where the extended action is obtained by introducing sources
$J_{\varphi^i}$ for all the fields, i.e.\
$S_{\rm ext}=S+\int J_{\varphi^i}\varphi^i$.
The non-invariance of the measure of this path
integral, together with the variation of the
renormalized extended action leads to an expression of
the anomalous Ward identity.

Under the symmetries (4) the partition function transforms as
follows:
\begin{equation}
Z\rightarrow Z^\prime=\int{\cal D}\varphi^i\exp \big(S_{\rm ext}
+\int J_{\varphi^i}\delta\varphi^i+\triangle\big),
\end{equation}
where the anomaly $\triangle$ includes a contribution ${\cal A}$ from
varying the measure and a contribution $\delta S$ from varying the
renormalised action. Expanding
the exponential and using the following Legendre transform:
\begin{equation}
\Gamma={\rm ln} Z-\int J_{\varphi^i}\varphi^i,
\end{equation}
we obtain the Ward identity in terms of the effective
action $\Gamma$. Explicitly, one has the relation
\begin{equation}
{\delta\Gamma\over\delta
K^{\varphi^i}}{\delta\Gamma\over\delta\varphi^i}=\triangle\cdot\Gamma,
\end{equation}
where $\triangle\cdot\Gamma$ denotes all one-particle-irreducible
graphs with precisely one insertion of
the composite anomaly operator $\triangle$. The local chiral $W_3$ anomalies
are already known from the explicit Feynman-diagram calculation of the
left-hand side of expression (13) \cite{mohayaeestelle}. In the
following, we shall derive them by an alternative path-integral method.

In evaluating $\triangle$, we first evaluate the terms ${\cal A}$,
steming from the measure.
Under the symmetries (4), the path-integral measure transforms
through a Jacobian factor which,
when properly regularized, leads to expressions that contribute to the
anomalies. The part of the Jacobian ${\cal
J}_1$ which contributes to the anomalies at order $\hbar$
is given by:
\begin{equation}
{\cal J}_1=\exp {\rm Tr} \left[{\partial\delta\varphi^i\over
\partial\varphi^j}\right]\bigg|_0=e^{{\cal A}_1},
\end{equation}
where the subscript ``$0$'' indicates terms of zeroth order in
$\hbar$ and the trace includes an integration over space-time.

The above Jacobian is singular, as it consists of products of delta
functions. These delta-functions arise both through taking
derivatives as well as through taking the subsequent trace
 in the above expression. This divergent Jacobian can, however,
be regularized  to give well-defined expressions which contribute to the local
anomalies at one-loop order.\footnote{
It is  well-known how to construct
regulators which lead to consistent
anomalies (i.e.\ anomalies that satisfy the Wess-Zumino consistency
condition). For example, one may use the Pauli-Villars
scheme \cite{diaz}.}
The regularized expression is:
\begin{eqnarray}
{\cal A}_1&=&\lim_{M\to\infty}{\rm Tr}\bigg(
\left[{\partial\delta\phi\over\partial\phi}\right]\bigg|_0e^{-H_\phi/M^2}+
\left[{\partial\delta
c\over\partial c}\right]\bigg|_0e^{-H_c/M^2}\nonumber\\
&+&\left[{\partial\delta\gamma\over\partial\gamma}\right]\bigg|_0
 e^{-H_\gamma/M^2}+
\left[{\partial\delta\pi_c\over\partial\pi_c}\right]\bigg|_0 e^{-H_{\pi_c}/M^2}
+\left[{\partial\delta\pi_\gamma\over\partial\pi_\gamma}\right]\bigg|_0
 e^{-H_{\pi_\gamma}/M^2}\bigg)\nonumber\\
& &
\end{eqnarray}
where the delta-functions that appear when evaluating the square
brackets are replaced by Gaussian regulators
\begin{equation}
H_\phi={\partial^2
S\over\partial\phi^i\partial\phi^j}
\end{equation}
for the scalar fields and
\begin{equation}
H_c=\bigg[\left({\partial^2
S\over\partial c\partial\pi_c}\right)^\dagger
{\partial^2 S\over\partial c\partial\pi_c}\bigg]\bigg|_{\rm chiral}
\end{equation}
for the spin-2 ghost and likewise for the rest of
 the fields \cite{diaz}\cite{ceresole}.
In the above expressions, the subscript ``chiral'' means we are
restricting to
terms chiral in gauge fields.

Evaluating the trace in (15) in the plane-wave basis, isolating the
infinite part of the resulting expression and also
calculating the contribution arising from the
variation of the action itself,  one is left
with the following expression for the anomaly at order $\hbar$:
\begin{eqnarray}
\triangle_1&=&{\cal A}_1+(\delta S)_1\nonumber\\
&=&\hbar
({16\over 30 \pi } (1-17a)-{a\over 12 \pi } C_{\rm mat}  )
\int d^2z\,  \Bigg(
{\gamma }\,  {{\pi}_c} \, \partial {\gamma }\,
( {\partial }^3 h - {\partial }^3
{K_{{\pi}_c}} ) \nonumber\\
& & +  {\partial }^3 c\,
[ {\gamma }\,  K_c \, \partial \gamma - {\pi}_c \, ( \partial \gamma
\, B - \gamma
\, \partial B - \partial \gamma\,  K_{{\pi}_{\gamma}} + \gamma \, \partial
K_{{\pi}_{\gamma}} ) ] \Bigg) \nonumber\\
& & - {\hbar \over 12 \pi } (100- C_{\rm mat}) \int d^2z\,
c\, ( {\partial }^3 h - {\partial }^3
{K_{{\pi}_c}}).
\end{eqnarray}

In the same manner,  we can evaluate the anomalies at order $\hbar^2$
arising from the variation of the renormalised action and also
from the Jacobian for the measure of the path integral.
Putting all terms together,  we obtain:
\begin{eqnarray}
\triangle_2^\prime &=&{\cal A}_2+(\delta S)_2\nonumber\\
&=& \hbar^2 \bigg( - {29\over 50 \pi } (1-17a) +
{C_{\rm mat}-2d_{ijk}d_{ijk}\over 360 \pi}\bigg)
\int d^2z\,  \gamma\, (\partial^5 B - \partial^5
K_{\pi_\gamma} ).
\end{eqnarray}
In evaluating the preceding order-$\hbar$ and -$\hbar^2$ anomalies,
we have found all the local anomaly contributions that are accessible
from knowledge of the
one-loop result.\footnote{Note that the prime on $\triangle^\prime_2$
in eqn.(19) indicates that since this contribution is obtained using
only
one-loop method, it might yet be accompanied by genuine two-loop
contributions,
as we shall see.}

Nevertheless, there is still further information about the anomalies
that can be deduced
from the above. Note that in expression (18) there are `off-diagonal'
terms mixing the spin-2 and spin-3 sectors. The presence of such
off-diagonal anomalies is particularly significant for the structure
of the anomalies at the
next order in $\hbar$, since further contractions are possible. In fact, the
dressing of these off-diagonal terms in
${\triangle_1}$ produces the following non-local anomaly contribution,
 ${\sl A}_{2, {\rm nl}}$,
at order ${\hbar^2}$:
\begin{eqnarray}
{\sl A}_{2, {\rm nl}}=\triangle_1\cdot S_0&=&\hbar^2
\bigg({16\over 30\pi}(1-17a)
-{a\over 12\pi}C_{\rm mat}\bigg)\nonumber\\
& &
\ \times\int d^2z\bigg[(\partial^3 h-\partial^3 K_{\pi_c})
\bigg(2\partial\gamma{\partial^2\over\bar\partial}
-{5\over 6}\gamma{\partial^3\over\bar\partial}
\bigg)(K_{\pi_\gamma}-B)\nonumber\\
& &
+\partial^3 c\bigg({5\over 6}(B-K_{\pi_\gamma})
{\partial^3\over\bar\partial}-
2(\partial B-\partial K_{\pi_\gamma})
{\partial^2\over\bar\partial}\bigg)(K_{\pi_\gamma}-B)\nonumber\\
& &-{3\over 10}\bigg(\partial\gamma(B-K_{\pi_\gamma})-\gamma(\partial B
-\partial
K_{\pi_\gamma})\bigg){\partial^5\over\bar\partial}(K_{\pi_c}-h)\bigg],
\end{eqnarray}
where $S_0$ is the classical part of the renormalised action (3). We have
now evaluated both the local and non-local parts of the
one-loop
order $W_{3}$ anomalies, to order $\hbar^2$.

\section{Consistency of the $W_3$ anomalies}
\indent

In the preceding section we derived one-loop anomalies for chiral $W_{3}$
gravity in a derivative gauge. We now wish to
consider the consistency of these derived anomalies and begin by
looking at the Ward identity.

At order $\hbar$ the Ward identity (13) reads
\begin{eqnarray}
\triangle_1&=&{\delta S_0 \over \delta
{\varphi }^i}\,  {\delta \Gamma_1 \over \delta K_{{\varphi }^i}}
+{\delta S_0 \over \delta K_{{\varphi }^i}}\,
{\delta \Gamma_1 \over \delta {\varphi }^i}
+{\delta S_{1\over 2 } \over \delta {\varphi }^i}\,
{\delta S_{1\over 2} \over \delta K_{{\varphi }^i}}\nonumber\\
&=& (S_0 ,\Gamma_1 ) + {1\over 2}(S_{1\over 2} ,S_{1\over 2}),
\end{eqnarray}
where the second line is written in the more compact anti-bracket
notation, defined by
\begin{equation}
(M,N)\equiv{\delta M \over \delta
{\varphi }^i}\,  {\delta N \over \delta K_{{\varphi }^i}}
+{\delta M \over \delta K_{{\varphi }^i}}\,
{\delta N \over \delta {\varphi }^i}.
\end{equation}
In order to obtain the well-known order $\hbar$ Wess-Zumino consistency
condition \cite{wess}, we take the anti-bracket of $S_0$ with the above
expression.
Simple manipulations using the Jacobi identity for the
anti-bracket, together with following low-order relations
\begin{equation}
(S_0, S_0)=0, \ \ \ \ \\ (S_0, S_{1/2})=0,
\end{equation}
lead to the desired result, namely:
\begin{equation}
(S_0, \triangle_1)=0.
\end{equation}
If we insert the order $\hbar$ anomaly derived in the previous section
(eqn.\ 18), a straightforward calculation verifies that it does indeed satisfy
this consistency condition to this order.

Next, we move on to consider the consistency of the anomalies at order
$\hbar^2$. To obtain the order $\hbar^2$ Wess-Zumino consistency
condition, we once again begin with the Ward identity (13) at the appropriate
order:
\begin{equation}
(S_0, \Gamma_2)+(S_{1/2}, \Gamma_{3/2})+{1\over
2}(\Gamma_1, \Gamma_1)=\triangle_2+{\sl A}_{2,{\rm nl}}.
\end{equation}
The consistency condition is then derived
as in the order $\hbar$ case,  noting also now that the local
contribution to the
anomaly at order $\hbar^{3/2}$ is zero. The resulting local condition
is given by
\begin{equation}
(S_0, \triangle_2)+(S_0, {\sl A}_{2, {\rm nl}})_{\rm loc}+
(S_1, \triangle_1)=0,
\end{equation}
where the second term is restricted to local contributions after
taking the antibracket.\footnote{Note
that the only non-local terms that can contribute to this
local Wess-Zumino consistency condition, are those that are
of first-order in non-locality
(i.e.\ with just one factor of $1/\bar\partial$).}
Any non-local condition would have to hold
separately.

On insertion of the calculated results (18, 19, 20) and
after a relatively tedious
calculation, we discover that this condition
only holds if there is an extra contribution to the local
order-$\hbar^2$ anomaly. This
contribution is given by
\begin{equation}
\triangle _2-\triangle_2^{\prime} =\hbar^2\bigg( {{2d_{ijk}d_{ijk}}
\over{ 360 \pi}}\bigg )
\int d^2z\,  {\gamma}\, ({\partial }^5 B - {\partial }^5
{K_{{\pi}_c}} )
\end{equation}
and stems from a genuine two-loop diagram.\footnote{
This approach to overcoming the deficiences of the Fujikawa method
has also been used in $W_\infty$ \cite{mohayaee}.} Recall that
the techniques employed in evaluating the anomalies in the
preceding section are necessarily restricted to the
one-loop order. Thus, having used the consistency conditions to overcome
the deficiencies of the Fujikawa method, we find that the anomalies
now agree with those already known from explicit Feynman-diagram
calculations \cite{mohayaeestelle}.

\section{Lam's theorem and non-local anomalies}
\indent

So far, we have derived the non-local anomalies at order $\hbar^2$ simply
via the dressing of the local one-loop anomalies. In this section, we will go
on to derive these non-local anomalies directly from the left-hand
side of the Ward
identity (13). Comparison of the two results will then
form an explicit check of
Lam's theorem \cite{lam}, which states that they should be identical. There
has been to date no direct check of
this theorem in an anomalous system as
complex as that of chiral $W_3$ gravity.

In checking Lam's theorem we are looking
for the terms that make non-local contributions to the
left-hand side of the Ward identity (13), at order $\hbar^2$.
Note that, for our purposes, we need only consider terms that
are of first order in non-locality,
for comparison with ${\sl A}_{2, {\rm nl}}$ (20).

Since this is a somewhat tedious calculation, details
of the relevant diagrams together with their contributions are
contained in Appendix A.

It is, however, worth commenting on the following point. At order
$\hbar$, a simple ``buck-passing'' mechanism exists. As an illustration,
consider the non-local terms arising from the spin-2 gauge field
self-energy diagram, shown in Fig.1(a). These are precisely
cancelled by the purely non-local diagram, Fig.1(b), of higher order in the
spin-2 gauge field. What remains after this cancellation
is simply the well-known
Virasoro anomaly.

\begin{figure}[htb]
\begin{center}
\leavevmode
\epsfbox{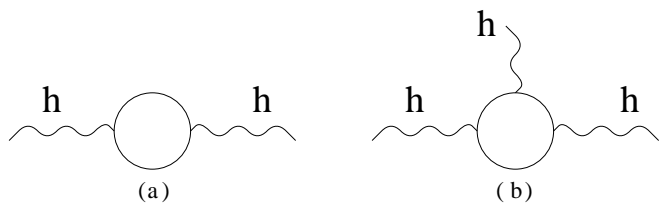}
\end{center}
\caption{``Buck-passing'' at order $\hbar$ \label{fig:spin2}}
\end{figure}

We might expect that this mechanism also applies at order $\hbar^2$.
If this were the case, the non-local terms, (which we must anticipate
if Lam's theorem is to hold), would have to arise entirely
from other types of diagrams.

At order $\hbar^2$, the illustration analogous to that above is to
consider the spin-3 gauge-field self-energy diagram shown in Fig.2,
\begin{figure}[htb]
\begin{center}
\leavevmode
\epsfbox{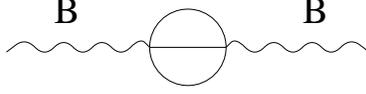}
\end{center}
\caption{Spin-3 self-energy diagram \label{fig:spin3}}
\end{figure}
since this is where the local anomaly arises. By comparison
with the order-$\hbar$ phenomenon, we might expect that the
non-local terms arising from this diagram
are precisely cancelled by contributions from
diagrams of higher order in the spin-2 gauge field.
In fact, taking into account
all loop orders that might feasibly give a cancellation
of the non-local terms resulting
from this diagram, (see Appendix A), we find that no such cancellation
occurs. No such ``buck-passing'' mechanism exists at order $\hbar^2$
and so no non-local term can be ignored.

The complete first-order non-local
contributions to the left-hand side of the Ward identity (13)
at order $\hbar^2$, having
examined all possible contributory diagrams, is given by:
\begin{eqnarray}
&&(\Gamma , \Gamma)_{\hbar^2 ,{\rm nl}}=\hbar^2\bigg({16(1-17a)\over
30\pi}-{aC_{\rm mat}\over 12\pi}\bigg)\nonumber\\
&\times&\int\bigg[\partial^3 c\bigg(-{5\over
6}(K_{\pi_\gamma}-B){\partial^3\over\bar\partial}+
2(\partial K_{\pi_\gamma}-\partial
B){\partial^2\over\bar\partial}\bigg)(K_{\pi_\gamma}-B)\nonumber\\
&+&\partial^3(K_{\pi_c}-h)\bigg({5\over 6}\gamma{\partial^3\over\bar\partial}
(K_{\pi_\gamma}-B)-2\partial\gamma{\partial^2\over\bar\partial}
(K_{\pi_\gamma}-B)\bigg)\bigg]\nonumber\\
&+&\bigg({16(1-17a)\over 30\pi}-{an\over
12\pi}-{10a\alpha_i\alpha_i\over 9\pi}\bigg)\nonumber\\
&\times&\int\bigg[{
3\over 10}\bigg(\partial\gamma(K_{\pi_\gamma}-B)-\gamma(\partial
K_{\pi_\gamma}-\partial
B)\bigg){\partial^5\over\bar\partial}(K_{\pi_c}-h)\bigg]\nonumber\\
&+&\int\bigg(-{n\over
360}+{d_{ijk}d_{ijk}-9e_{ij}e_{ij}-6e_{ij}e_{ji}\over
180}\bigg)\nonumber\\
&\times&{\partial^2\over\bar\partial}(K_{\pi_c}-h)\bigg(\partial\gamma\partial^3-\partial^3\gamma\partial
+2\partial^4\gamma-2\gamma\partial^4
\bigg)(K_{\pi_\gamma}-B)\nonumber\\
&-&\int{e_{ij}(e_{ij}+e_{ji})\over 12}\bigg(\gamma(\partial^4
K_{\pi_\gamma}-\partial^4
B)-\partial^4\gamma(K_{\pi_\gamma}-B)\bigg){\partial^2\over\bar\partial}
(K_{\pi_c}-h)\nonumber\\
&-&\int{d_{iij}f_j\over
6}{\partial^2\over\bar\partial}(K_{\pi_c}-h)\bigg(\partial\gamma\partial^3+\gamma\partial^4 -\partial^4
\gamma-\partial^3\gamma\partial\bigg)
(K_{\pi_\gamma}-B).\nonumber\\
\end{eqnarray}

After invoking the conditions (7), we indeed find that the above Ward
identity calculation yields the expected result
of ${\sl }A_{2, {\rm nl}}$. With this,  we
have verified that Lam's theorem holds for the non-trivial example
of fully quantised chiral $W_3$ gravity.

\section{Non-local anomalies from the BRST charge}
\indent

The failure of the BRST charge $Q$ to be nilpotent also
gives an expression for the anomaly. Whether we investigate the
anomalies of a theory through the BRST charge or the Ward identity,
compatible results should be expected. Since non-local anomalies are found
when considering the Ward
identity, such terms should also be found in the BRST charge approach.
However, as the BRST charge contains a contour integral,
it might naively be considered to be a purely local expression. Here, we
show how the non-local features do indeed appear in the BRST-charge
approach to the anomalies.

The full quantum level BRST charge in the derivative gauge is given by
\cite{mohayaeestelle}\cite{thierry-mieg}
\begin{equation}
Q=\oint c(T_{\rm mat}+ {1\over 2}T_{\rm gh})+\gamma(W_{\rm
mat}+{1\over 2}W_{\rm
gh})+\pi_h\pi_b+\pi_B\pi_\beta .
\end{equation}

If we interpret the BRST-charge
as the integral of a normal-ordered operator current:
\begin{eqnarray}
Q=\oint {dz \over 2\pi i}J(z),
\end{eqnarray}
where the integral in complex world-sheet coordinates is
taken to be a closed loop around the origin\footnote{Since
Cauchy's theorem is used to collect simple poles,  the factor
of $(2\pi i)^{-1}$ appears in the measure so as to make
contact with the standard formulae.}
,  then we may use
standard operator-product techniques to obtain the expectation value
of ${Q}^2$. To do this, one would consider correlators such as
\begin{eqnarray}
\left<Q^2\cdots\right>=\int{\cal D}\varphi^i (w)
\oint {dz \over 2\pi i} {<JJ>}_1(z) \exp S(w)\cdots,
\end{eqnarray}
where $\varphi^i$ stands generically for all the fields that are
contained in the BRST action (3),
the subscript ``1'' in ${<JJ>}_1$ denotes taking the residue
 of the first-order pole $(z-w)^{-1}$, in the Laurent series resulting
from the operator-product $J(z)J(w)$ and $\cdots$ stands for any other
operator insertion.

One may replace correlator relations like (31) by
equivalent operator relations. Expanding the exponential,  the known
expression \cite{mohayaeestelle} for the local
part of $Q^2$ arises from the first term
\begin{eqnarray}
Q^2_{\rm loc}&=&\oint{(100-C_{\rm mat})\over 6}c\partial^3 c+\bigg(-{16\over
15}(1-17a)+{a\over 6}C_{\rm
mat}\bigg)\gamma\pi_c\partial\gamma\partial^3 c\nonumber\\
& &+\hbar \bigg({29\over
25}(1-17a)-{C_{\rm mat}\over
180}\bigg)\gamma\partial^5\gamma,
\end{eqnarray}
whilst the following terms, of first order in non-locality,  stem
from the second
\begin{eqnarray}
Q^2_{\rm nl}&=&\oint {dz \over 2\pi i}\big< {<JJ>}_1 (z),
S\big>\nonumber\\
&=&\oint \hbar \bigg(-{16(1-17a)\over 15}+{a\over 6} C_{\rm mat}
\bigg)\bigg({3\over 10}\gamma\partial\gamma{\partial^5\over
\bar\partial}(K_{\pi_c}-h)\nonumber\\
& &-2\partial^3
c\partial\gamma{\partial^2\over\bar\partial}(K_{\pi_\gamma}-B)+{5\over
6}\partial^3
c\gamma{\partial^3\over\bar\partial}(K_{\pi_\gamma}-B)\bigg).
\end{eqnarray}

Thus we see that non-local terms arise similarly in the BRST approach
and have their roots in the off-diagonal
part of the local anomaly.

\section{Relating the BRST algebra anomaly to the BRST
Ward identity anomaly}
\indent

In ref.\cite{mohayaeestelle}, a relationship was proposed
between the Ward-identity anomalies and those obtained from
the square of the BRST charge.
It was shown that the local parts of the one- and two- loop
$W_3$ anomalies arising in the BRST Ward identity can be obtained
by the action of the anomalous operator $Q^2$ on the gauge fermion.
Explicitly, one has the relation
\begin{equation}
\Delta_1 + \Delta_2 = -{1\over 2\pi}\int d^2 v\bigg< <JJ>_1 , \Psi (v)\bigg>_1.
\end{equation}

In this section, we explore whether or not this relationship can be
extended to include the non-local anomalies.

{}From the action in ``canonical BRST'' form
$\pi^a\bar\partial q^a-\delta\Psi$ (3)  we can see that the gauge
fermion $\Psi$ for our action (3) is given by:
\begin {equation}
\Psi=\pi_c h+\pi_{\gamma} B + \sum_{i} (-1)^{[i]} \varphi^i K_{\varphi^i},
\end{equation}
where $[i]$ takes the values (0, 1) for (bose, fermi) variables.
Taking the operator product of the non-local part of $Q^2$
with this gauge fermion,  relatively straightforward algebra
verifies that:
\begin{equation}
A_{2, {\rm nl}} =  -{1\over 2\pi}\int d^2 v
\bigg< \big<<JJ>_1, S\big>, \Psi (v)\bigg>_1.
\end{equation}
Thus,  the non-local Ward-identity anomalies
can indeed be obtained from the action of the non-local
part of $Q^2$ acting on the gauge fermion.

\section{Gauge-independence of the anomaly relation}
\indent

In the preceding section we saw that the relationship between the
BRST algebra anomalies (32, 33) and those in the BRST Ward identity
(18, 19, 20, 27), originally proposed for local
anomalies, also holds
for the non-local terms. In this section, we investigate the gauge
dependence of this anomaly relation by choosing a more general
gauge-fixing condition.

Let us consider chiral $W_3$ gravity in a parametrised gauge,
specifically $\zeta\bar\partial h+\mu h=0$ and for the spin-3 sector
 $\zeta^\prime\bar\partial B+\mu^\prime B=0$. The action in this gauge
is
\begin{equation}
S={1\over\pi}\int \bigg[-{1\over 2}
\bar\partial\phi^i \partial\phi^i+\pi_h\bar\partial h+\pi_B\bar\partial
B-\pi_b\bar\partial b-\pi_c\bar\partial
c-\pi_\beta\bar\partial\beta-\pi_\gamma\bar\partial\gamma
-\delta \Psi\bigg],
\end{equation}
where the gauge fermion $\Psi$ is $\left(h\pi_c-\mu h
b+B\pi_\gamma-\mu^\prime B\beta+\sum_i(-1)^{[i]}\varphi^i
K_{\varphi^i}\right)$ and the variations are as in (4) apart from
\begin{equation}
\delta h={\pi_b\over\zeta}\,\,\,;\,\,\,\delta B={\pi_\beta\over\zeta^\prime}.
\end{equation}

The Ward-identity anomalies remain as in the derivative
gauge (18, 19, 20, 27),
since the new terms arising in the action (37) in this parametrised
gauge make no additional contribution.
The BRST charge similarly remains unchanged from the derivative gauge
expression (29).

If we note the following modified operator-product expansions:
\begin{equation}
\pi_h(z)h(w)\sim{\hbar\over\zeta(z-w)}\,\,\,;\,\,\,\pi_B(z)B(w)
\sim{\hbar\over\zeta^\prime(z-w)},
\end{equation}
then it is simple to check that the anomaly relations (34, 36) hold
independently of the gauge choice.

It is worth emphasising a curious feature of this result: in a theory
with gauge anomalies, the anomalous operator $Q^2$ annihilates a certain
part of the gauge fermion $\Psi$ to give a gauge-independent result
despite the fact that it is $\Psi$ that establishes the choice of gauge.

\section{General validity of the Anomaly Relation}
\indent

So far we have seen that the anomaly relation holds for local and
non-local chiral $W_3$ anomalies, under a general gauge-fixing condition.
This motivates seeking an underlying reason for
this relation, independent of the
theory concerned. In this section we give a
heuristic argument for the anomaly relation.

In conventional field theory the anomaly is explored through
the Ward identity. To obtain a mathematical expression for the Ward
identity we can take either of two approaches. The Ward identity may
be expressed in terms of a classically-conserved Noether
current or in terms of the effective action after introducing
sources into the theory (see Section 2).

In either case, one starts with the partition function $Z$
and checks that it is invariant under the relevant field
transformations. In the first approach, the Ward identity is obtained by
considering the variation of the partition function under a change of
integration variables of the form of classical symmetry transformation,
but with a space-time dependent parameter $\alpha$, as follows:
\begin{equation}
Z\longrightarrow Z^\prime=\int {\cal D}\varphi^i \exp(
\int\big({\cal L}+K_{\phi^i}\delta\phi^i\big)
-\int\alpha\partial_\mu J^\mu+\alpha{\sl A}),
\end{equation}
where ${\sl A}$ is the anomaly,
 $\alpha\partial_\mu J^\mu$ is the variation of the
Lagrangian, $J^\mu$ being the Noether current and $K_{\phi^i}$ are the
sources for the variations of the fields $\phi^i$. Expanding
the last two terms in
the exponent and requiring the invariance of the partition function
under changes of integration variables,
we obtain
\begin{equation}
\int\partial_\mu J^\mu={\sl A}.
\end{equation}
This is a statement of the Ward identity in terms of the Noether
current.

In the second approach, we introduce sources $J_{\varphi^i}$ for the
fields $\varphi^i$. The
variation of the
partition function, under a global symmetry,
 in terms of the extended Lagrangian ${\cal L}
_{\rm ext}={\cal L}+J_{\varphi^i}\varphi^i+K_{\varphi^i}\delta\varphi^i$
is given by
\begin{equation}
Z\longrightarrow Z^\prime=\int {\cal D}\varphi^i \exp(\int{\cal
L}_{\rm ext}
+\int J_{\varphi^i}\delta\varphi^i+{\sl A}).
\end{equation}
Expanding the last two terms in this exponent, and using a Legendre
transformation (12), to obtain the effective action $\Gamma$ from the
partition function, we find the well-known expression of the Ward identity,
\begin{equation}
{\delta\Gamma\over\delta\varphi^i}{\delta\Gamma\over\delta
K_{\varphi^i}}
={\sl A}.
\end{equation}
 Hence, we can see that the Ward identity
anomaly can equally well be expressed either way.

Now, we can use this information to consider the general validity of
our anomaly relation. Since the BRST charge $Q$ is the spatial
integral of the time component of the Noether current $J$, we have
\begin{equation}
{\sl A}={\delta\Gamma\over\delta\varphi^i}{\delta\Gamma\over\delta
K_{\varphi^i}}={\partial Q\over\partial t}=-[Q, H],
\end{equation}
where $H$ is the Hamiltonian.
In all reparametrisation-invariant theories written in
``canonical''
form similar to (3) the Hamiltonian can be
expressed as the variation of a gauge
fermion $\Psi$ \cite{gomisparissamuel}, hence,
\begin{equation}
{\sl A}=-[Q, \{Q, \Psi\}].
\end{equation}
Subsequent use of the Jacobi identity in the right hand side then
yields,
\begin{equation}
{\sl A}= -{1\over 2}\bigg[\{Q, Q\}, \Psi\bigg].
\end{equation}

The above relation is the same as that discussed in
Sections 7 and 8, although there we
used operator-product language rather than commutators.

\section{Conclusion and Outlook}
\indent

In this paper, we have made a detailed survey of the structure of
$W_3$ anomalies. In our work, we used the
Romans' realisation for a fully quantised $W_3$ algebra.
Using the Romans' renormalisation
conditions ensures exactly the cancellation of matter-dependent anomalies.
We investigated the remaining anomalies up to order $\hbar^2$,
using several different approaches.

Firstly, the Fujikawa method of anomaly derivation was used to find the local
order $\hbar$ and $\hbar^2$ anomaly contributions. Non-local order
$\hbar^2$ terms were then found by dressing the off-diagonal order
$\hbar$ anomaly. The Fujikawa path-integral approach is a purely
one-loop technique. We showed that genuine two-loop contributions
can nevertheless be accessed by demanding that the Wess-Zumino consistency
conditions hold. It remains an interesting problem to discover
how the two-loop anomalies may be handled directly in the path-integral
formalism. This might be an equivalent problem to that of finding how the
Pauli-Villars regularisation can be extended to the regularisation of
diagrams beyond one-loop.

Next, the non-local anomalies at order $\hbar^2$ were calculated again,
this time directly from Feynman diagrams. It was shown that
these non-local contributions to the Ward-identity precisely match
those from the dressing of the one-loop anomaly. This constitutes
a non-trivial check of Lam's theorem. We also
noted that the order-$\hbar$ ``buck-passing'' mechanism, which is usually
taken for granted at order $\hbar$,
does not apply at order $\hbar^2$.
Explicitly, the non-local anomaly contributions from the
spin-3 gauge-field self-energy diagram are not simply
cancelled by terms from diagrams of higher order in the gauge
fields.

We further considered the
$Q^2$ anomaly for the BRST algebra. Although it is well known
how to obtain local terms this way, non-local
anomalies have rarely been discussed. Thus, we looked closely
at the expectation value of the square of the BRST charge as a means of
calculating all the anomalies. Again,
both the local and non-local contributions were found up to order
$\hbar^2$.

We also discussed the relationship
between the anomalies found in the two approaches. The anomaly
relation proposed in ref.\cite{mohayaeestelle} requires that the BRST
Ward identity anomalies
be proportional to the application
of the square of the BRST charge to the gauge fermion. This relation,
previously discussed for local anomaly contributions in chiral
gravities, was shown to
apply not only to the non-local terms but also to hold under a more
general gauge-fixing condition and furthermore to be valid
for all reparametrisation-invariant theories.
It remains an open problem to see if the anomaly relation holds
for a more general class of theories, for example Yang-Mills theory.

\section*{Acknowledgment}
\indent

We are indebted to Kelly Stelle for suggesting the problem, continuing
advice and for critical reading of the manuscript. We would also like
to thank Luis Garay, Jordi Paris, Kostas Skenderis, Kris Thielemans,
Arkady Tseytlin, Stefan Vandoren and
Toine Van Proeyen for helpful discussions.

\section*{Appendix A}
\indent

In the following we use the Ward identity (13) to evaluate the non-local
anomalies at order $\hbar^2$.

For simplicity,  we list only the possible contributions that would
result in terms with no explicit factors of $\phi^i , b, \beta$,  any of
the momenta or any of the sources except for $ K_{\pi_{c}},
K_{\pi_{\gamma}}$. They should also be of no more
than second order in the gauge fields.
This is to facilitate comparison with the result obtained via the
dressing of the one-loop anomaly (20).

The following operator-product expansions obtained from the
action (3) \cite{thielemans}
are used in the evaluation of the diagrams presented in this appendix:
\begin{eqnarray}
\partial {\phi }^i (z) \partial {\phi }^i (w) &\sim& {-
\hbar\over (z-w)^2},\nonumber\\
{\pi }_h (z) h (w) &\sim&
{\hbar\over z-w };\,\,\,\,\,\,\,\,
{\pi }_B (z) B (w) \sim {\hbar\over z-w },\nonumber\\
 c(z){\pi }_c (w) &\sim& {\hbar \over {z-w}};\,\,\,\,\,\,\,\,
b(z){\pi }_b (w)\sim {\hbar\over
{z-w}},\nonumber\\
\gamma(z){\pi }_{\gamma } (w) &\sim& {\hbar \over {z-w}};\,\,\,\,\,\,\,\,
\beta (z){\pi
}_{\beta } (w) \sim {\hbar\over {z-w}}.
\end{eqnarray}
Let us firstly consider the simple loop:
\begin{figure}[hb]
\begin{center}
\leavevmode
\epsfbox{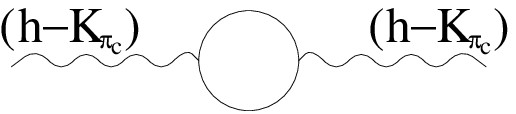}
\end{center}
\end{figure}

\noindent
which contributes as a $\phi$ loop,a spin-2 ghost loop and a spin-3 ghost
loop, giving:
\begin{eqnarray}
{\delta\Gamma_{1}\over\delta K_{\pi_c}}{\delta S_{1}\over\delta
\pi_c}&=&-{\hbar^2(100-n)(1-17a)\over
360\pi}\nonumber\\
& &
\times\int\bigg[2\partial^3\gamma(K_{\pi_\gamma}-B)-3\partial^2\gamma
(\partial K_{\pi_\gamma}-\partial
B )\nonumber\\
& &
+3\partial\gamma(\partial^2K_{\pi_\gamma}-\partial^2
B)-2\gamma(\partial^3K_{\pi_\gamma}-\partial^3
B)\bigg]{\partial^3\over\bar\partial}(K_{\pi_c}-h).\nonumber\\
& &
\end{eqnarray}

Next, we consider contributory diagrams with external $\phi$ lines.
The first in this series of diagrams is the following $\phi$ loop:
\begin{figure}[htb]
\begin{center}
\leavevmode
\epsfbox{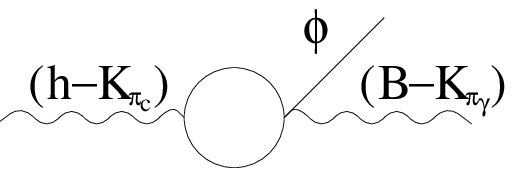}
\end{center}
\end{figure}

\indent
which contributes:
\begin{equation}
{\delta\Gamma_{1}\over\delta\phi^{i}}{\delta
S_{1}\over\delta K_{\phi^{i}}}=-{\hbar^2 d_{iij}f_j\over
6\pi}\int(K_{\pi_\gamma}-B)\partial^3\gamma
{\partial^3\over\bar\partial}(K_{\pi_c}-h).
\end{equation}

\vfill\eject
The second is the mixed spin-2 and spin-3 ghost loop:
\begin{figure}[htb]
\begin{center}
\leavevmode
\epsfbox{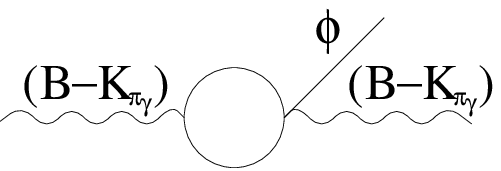}
\end{center}
\end{figure}

\indent
which contributes:
\begin{eqnarray}
& &
{\delta\Gamma_{3/2}\over\delta\phi^{i}}
\delta\phi^{i}_{1/2}={\hbar^2
a\alpha_i\alpha_i\over\pi}\int\partial^3c\bigg[{5\over 6}
(K_{\pi_\gamma}-B){\partial^3\over\bar\partial}-2(\partial
K_{\pi_\gamma}-\partial
B){\partial^2\over\bar\partial}\bigg](K_{\pi_\gamma}-B).\nonumber\\
  & &
\end{eqnarray}
The final external $\phi$ contribution is the $\phi$ loop corresponding
to the above diagram, giving:
\begin{equation}
{\delta\Gamma_{3/2}\over\delta\phi^{i}}\delta\phi^{i}_{1/2}=
-{\hbar^2e_{ij}d_{ijk}\alpha_k\over
6\pi}\int\partial^2c(K_{\pi_\gamma}-B)
{\partial^4\over\bar\partial}(K_{\pi_\gamma}-B).
\end{equation}

The next set of diagrams have external $\gamma$ lines.
The first is the $\phi$ loop:
\begin{figure}[htb]
\begin{center}
\leavevmode
\epsfbox{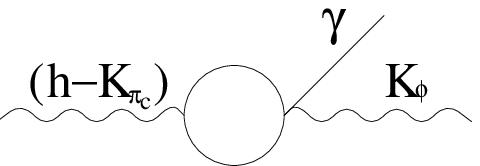}
\end{center}
\end{figure}

\noindent
which contributes:
\begin{equation}
{\delta\Gamma_{1}\over\delta K_{\phi^{i}}}{\delta
S_{1}\over\delta\phi^{i}}={\hbar^2d_{iij}f_j\over
6\pi}\int\gamma(\partial^3K_{\pi_{\gamma}}
-\partial^3B){\partial^3\over\bar\partial}(K_{\pi_c}-h).
\end{equation}
The mixed spin-2 ghost, $\phi$ loop:
\begin{figure}[hbt]
\begin{center}
\leavevmode
\epsfbox{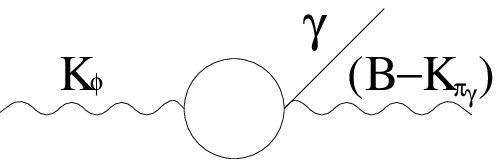}
\end{center}
\end{figure}

\noindent
gives:
\begin{equation}
{\delta\Gamma_{3/2}\over\delta K_{\phi^{i}}}{\delta
S_{1/2}\over\delta\phi^{i}}=
-{\hbar^2a\alpha_i\alpha_i\over\pi}\int\bigg[\partial
\gamma(K_{\pi_\gamma}-B)-\gamma(\partial
K_{\pi_{\gamma}}-\partial
B)\bigg]{\partial^5\over\bar\partial}(K_{\pi_c}-h).
\end{equation}

\vfill\eject
\noindent
The mixed spin-2, spin-3 ghost loop:
\begin{figure}[hbt]
\begin{center}
\leavevmode
\epsfbox{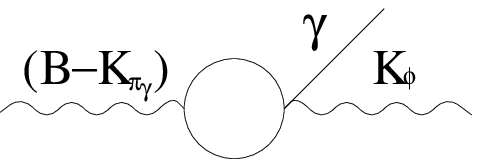}
\end{center}
\end{figure}

\noindent
contributes:
\begin{equation}
{\delta\Gamma_{3/2}\over\delta K_{\phi^{i}}}{\delta
S_{1/2}\over\delta\phi^{i}}=
-{\hbar^2a\alpha_i\alpha_i\over\pi}\int\bigg[2\partial\gamma{\partial^2\over\bar\partial}
(K_{\pi_\gamma}-B)-{5\over
6}\gamma{\partial^3\over\bar\partial}(K_{\pi_\gamma}-B)\bigg](\partial^3
K_{\pi_c}-\partial^3 h).
\end{equation}
The next contribution is from the $\phi$ loop corresponding to the
previous diagram, which gives:
\begin{equation}
{\delta\Gamma_{3/2}\over\delta K_{\phi^{i}}}{\delta
S_{1/2}\over\delta\phi^{i}}={\hbar^2e_{ij}d_{ijk}\alpha_k\over
6\pi}\int\gamma(\partial^2K_{\pi_c}-\partial^2
h){\partial^4\over\bar\partial}(K_{\pi_\gamma}-B).
\end{equation}
In the following figures (a) is a $\phi$ loop and (b) has
contributions from a mixed spin-2, spin-3 ghost loop as well as
a pure $\phi$ loop.
\begin{figure}[htb]
\begin{center}
\leavevmode
\epsfbox{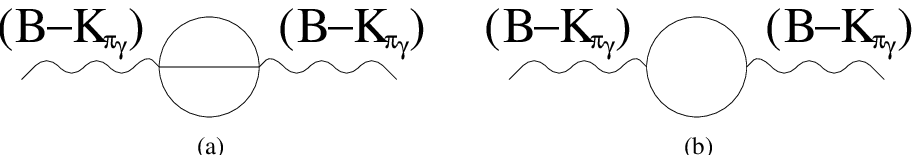}
\end{center}
\end{figure}

\noindent
Together they contribute:
\begin{eqnarray}
{\delta\Gamma_{2}\over\delta K_{\pi_\gamma}}{\delta
S_{0}\over\delta\pi_\gamma}&=&\hbar^2\bigg({-29(1-17a)\over
50\pi}+{(d_{ijk}d_{ijk}-9e_{ij}e_{ij}-6 e_{ij} e_{ji})\over
180\pi}\bigg)\int{\partial^5\over\bar\partial}(K_{\pi_\gamma}-B)\nonumber\\
& &\times\bigg[2(\partial K_{\pi_c}-\partial
h)\gamma-(K_{\pi_c}-h)\partial\gamma+(\partial K_{\pi_\gamma}-\partial
B)c-2(K_{\pi_\gamma}-B)\partial c\bigg].\nonumber\\
& &
\end{eqnarray}
This concludes the external $\gamma$ diagrams.

The bow-tie diagram,
\begin{figure}[htb]
\begin{center}
\leavevmode
\epsfbox{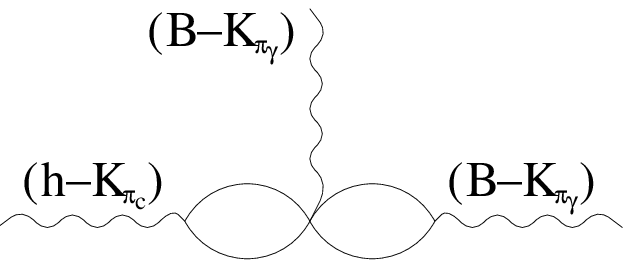}
\end{center}
\end{figure}

\noindent
with one $\phi$ loop
and one mixed spin-2, spin-3 ghost loop, contributes:
\begin{eqnarray}
& &
{\delta\Gamma_{2}\over\delta K_{\pi_c}}{\delta S_{0}\over\delta\pi_c}+
{\delta\Gamma_{2}\over\delta K_{\pi_\gamma}}{\delta
S_{0}\over\delta\pi_\gamma}=\nonumber\\
& &
-{\hbar^2an\over 12\pi}\bigg\{\int\partial^3c\bigg[2(\partial
K_{\pi_\gamma}-\partial B){\partial^2\over\bar\partial}(K_{\pi_\gamma}-B)
-{5\over 6}(K_{\pi_\gamma}-
B){\partial^3\over\bar\partial}(K_{\pi_\gamma}-B)\bigg]+
\nonumber\\
& &
\bigg[{5\over
6}\gamma{\partial^3\over\bar\partial}
(K_{\pi_\gamma}-B)-2\partial\gamma{\partial^2\over\bar\partial}
(K_{\pi_\gamma}-B)\bigg](\partial^3K_{\pi_c}-\partial^3
h)+\nonumber\\
& &
\bigg[2(\partial K_{\pi_\gamma}-\partial
B)\partial^2\gamma-2\partial\gamma
(\partial^2K_{\pi_\gamma}-\partial^2
B)\nonumber\\
& &
-{5\over 6}(K_{\pi_\gamma}-B)\partial^3\gamma+{5\over
6}\gamma(\partial^3K_{\pi_\gamma}-\partial^3
B)\bigg]{\partial^3\over\bar\partial}(K_{\pi_c}-h)\bigg\}.
\end{eqnarray}
There are also contributions from some three point functions
besides the bow-tie diagram. Consider the three-vertex diagram:
\begin{figure}[htb]
\begin{center}
\leavevmode
\epsfbox{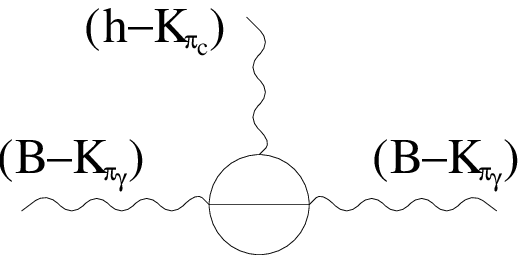}
\end{center}
\end{figure}

\noindent
This $\phi$ loop gives:
\begin{eqnarray}
& &{\delta\Gamma_{2}\over\delta K_{\pi_c}}{\delta S_{0}\over\delta\pi_c}+
{\delta\Gamma_{2}\over\delta K_{\pi_\gamma}}{\delta
S_{0}\over\delta\pi_\gamma}
=\nonumber\\
& &
-{\hbar^2d_{ijk}d_{ijk}\over 360\pi}\bigg\{\int\partial^5\gamma\bigg[2(\partial
K_{\pi_\gamma}-\partial
B){1\over\bar\partial}-4(K_{\pi_\gamma}-B)
{\partial\over\bar\partial}\bigg](K_{\pi_c}-h)+\nonumber\\
& &
(\partial^5 K_{\pi_\gamma}-\partial^5 B)
\bigg[4\gamma{\partial\over\bar\partial}-2
\partial\gamma{1\over\bar\partial}\bigg](K_{\pi_c}-h)+\nonumber\\
& &
{\partial^5\over\bar\partial}(K_{\pi_\gamma}-B)\bigg[2(\partial
K_{\pi_\gamma}-\partial B)c-4(K_{\pi_\gamma}-B)\partial
c+4\gamma(\partial K_{\pi_c}-\partial
h)-2\partial\gamma(K_{\pi_c}-h)\bigg]\bigg\}.\nonumber\\
& &
\end{eqnarray}
\vfill\eject
\noindent
There is also a three-point diagram generated by the counterterms:
\begin{figure}[htb]
\begin{center}
\leavevmode
\epsfbox{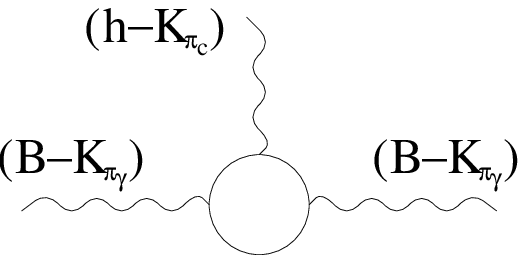}
\end{center}
\end{figure}

\noindent
This $\phi$ loop contributes:
\begin{eqnarray}
& &{\delta\Gamma_{2}\over\delta K_{\pi_\gamma}}{\delta
S_{0}\over\delta\pi_\gamma}+{\delta \Gamma_{2}\over\delta K_{\pi_c}}
{\delta S_{0}\over\delta\pi_c}=-{\hbar^2\big(e_{ij}e_{ij}+
2e_{ij}e_{ji}\big)\over
60\pi}\nonumber\\
& &
\times\int\bigg\{\bigg[\partial\gamma(K_{\pi_c}-h)+2(K_{\pi_\gamma}-B)\partial
c-c(\partial K_{\pi_\gamma}-\partial B)-2\gamma(\partial
K_{\pi_\gamma}-\partial
h)\bigg]{\partial^5\over\bar\partial}(K_{\pi_\gamma}-B)\nonumber\\
& &
+\bigg[\partial\gamma{1\over\bar\partial}
(K_{\pi_c}-h)-2\gamma{\partial\over\bar\partial}
(K_{\pi_c}-h)\bigg](\partial^5K_{\pi_c}-\partial^5
B)\nonumber\\
& &
+\bigg[2(K_{\pi_\gamma}-B){\partial\over\bar\partial}(K_{\pi_c}-h)-(\partial
K_{\pi-\gamma}-\partial B)
{1\over\bar\partial}(K_{\pi_c}-h)\bigg]\partial^5\gamma\bigg\}\nonumber\\
& &
-{\hbar^2(e_{ij}e_{ij}+e_{ij}e_{ji})\over
12\pi}\int\bigg[\gamma(\partial^2 K_{\pi_c}-\partial^2
h){\partial^4\over\bar\partial}(K_{\pi_\gamma}-B)\nonumber\\
& &
+\gamma{\partial^2\over\bar\partial}\partial^4(K_{\pi_\gamma}-B)-
(K_{\pi_\gamma}-B)\partial^2
c{\partial^4\over\bar\partial}(K_{\pi_\gamma}-B)\nonumber\\
& &
-(K_{\pi_\gamma}-B){\partial^2\over\bar\partial}
(K_{\pi_c}-h)\partial^4\gamma\bigg].
\end{eqnarray}
Finally, there are also ghost-loop contributions from
the preceding three-point diagram, contributing to
the anomaly as follows:
\begin{eqnarray}
& &
{\hbar^2(1-17a)\over 30\pi}\bigg[{-644\over
30}\bigg(\partial^6\gamma(K_{\pi_\gamma}-B)
+\gamma\partial^6(K_{\pi_\gamma}-B)\bigg)\nonumber\\
& &
-24\bigg(\partial^4\gamma\partial^2(K_{\pi_\gamma}-B)
+\partial^2\gamma\partial^4(K_{\pi_\gamma}-B)\bigg)\nonumber\\
& &
-{101\over
2}\bigg(\partial^5\gamma\partial(K_{\pi_\gamma}-B)
-\partial\gamma\partial^5(K_{\pi_\gamma}-B)\bigg)\bigg]
{1\over\bar\partial}(K_{\pi_c}-h)\nonumber\\
& &
+{\hbar^2(1-17a)\over 30\pi}(K_{\pi_c}-h)\bigg[{322\over
15}\gamma{\partial^6\over\bar\partial}
+{221\over
5}\partial\gamma{\partial^5\over\bar\partial}\nonumber\\
& &
+{744\over
9}\partial^3\gamma{\partial^3\over\bar\partial}
+56\partial^2\gamma{\partial^4\over\bar\partial}
+32\partial^4\gamma{\partial^2\over\bar\partial}
\bigg](K_{\pi_\gamma}-B)
\nonumber\\
& &+{\hbar^2(1-17a)\over 30\pi}\, c\, \bigg[{322\over
15}(K_{\pi_\gamma}-B){\partial^6\over\bar\partial}
+{221\over
5}\partial(K_{\pi_\gamma}-B){\partial^5\over\bar\partial}\nonumber\\
& &
+{744\over
9}\partial^3(K_{\pi_\gamma}-B){\partial^3\over\bar\partial}
+56\partial^2(K_{\pi_\gamma}-B){\partial^4\over\bar\partial}
+32\partial^4(K_{\pi_\gamma}-B){\partial^2\over\bar\partial}
\bigg](K_{\pi_\gamma}-B).\nonumber\\
& &
\end{eqnarray}

This concludes all the non-local order-$\hbar^2$ contributions that are
relevant for comparison with the result obtained from the
dressing of one-loop anomaly. This is indeed the complete result since all
other contributions cancel amongst themselves.

\end{document}